\documentclass[11pt]{article}
\usepackage[utf8]{inputenc}
\usepackage{yfonts}
\usepackage{comment}
\usepackage{answers}
\usepackage{tikz}
\usepackage{ytableau}
\usepackage{xcolor}
\usepackage{caption}
\usepackage{setspace}
\usepackage{graphicx}
\usepackage{enumitem}
\usepackage{multicol}
\usepackage{amsfonts}
\usepackage{mathrsfs}
\usepackage{wrapfig}
\usepackage{multicol}
\usepackage{cite}
\usepackage{hyperref}
\usepackage{url}
\usepackage[top = 2 cm, bottom = 2 cm, left = 2.5 cm, right = 1.5 cm]{geometry}
\usepackage{amsmath,amsthm,amssymb}

\definecolor{armygreen}{rgb}{0.29, 0.33, 0.13}

\graphicspath{{pictures/}}
\DeclareGraphicsExtensions{.pdf,.png,.jpg}

\newcommand{\Tr}{\operatorname{Tr}}

\begin{document}
\begin{flushright} \small
ITEP-TH-07/22\\
MIPT-TH-05/22
 \end{flushright}
\smallskip
\begin{center}
\begin{Large}\fontfamily{cmss}
\fontsize{15pt}{27pt}
\selectfont
	\textbf{Superintegrability in $\beta$-deformed Gaussian Hermitian matrix model from $W$-operators}
	\end{Large}
	
\bigskip \bigskip

\begin{large}
V. Mishnyakov$^{a,b,c,d},$\footnote{mishnyakovvv@gmail.com; }, A. Oreshina$^{a,d,}$ \footnote{oreshina.aa@phystech.edu; }
\end{large}

\bigskip

\begin{small}
$^a$ {\it MIPT, Dolgoprudny, 141701, Russia}\\
$^b$ {\it Lebedev Physics Institute, Moscow 119991, Russia}\\
$^c$ {\it ITMP MSU, Moscow 119991, Russia}\\
$^d$ {\it NRC "Kurchatov Institute", Moscow 117218, Russia}\\

\end{small}
 \end{center}

\bigskip
\begin{abstract}
    This paper is devoted to the phenomenon of superintegrability. This phenomenon is manifested in the existence of a formula for character averages, expressed through the same characters at special points and of its various generalization. In this paper we
    develop a method of proving such formulas from first principle from Virasoro constraints and $W$-representation. We apply it to prove the formula for the Jack functions averages - appropriate analogue of characters for the $\beta$-deformed Hermitian Gaussian matrix model. We also sketch the construction of $W$-operators from Calogero-Ruijsenaars Hamiltonians.

\end{abstract}
\section{Introduction}
		Recently an interesting property of matrix models called \emph{superintegrability} was brought to attention, see \cite{Mironov:2022fsr} for a rather extensive summary and references therein. It appears that in a wide variety of matrix models the are explicit formulae for averages of an appropriately chosen basis in the space of gauge invariant operators. In each situation these operators correspond to some symmetric polynomials. Usually, these polynomials correspond to characters of some algebra, like in the case of Schur functions or $Q$-functions  or some appropriate generalization of characters, like in this paper. Remarkably, expectation values of these polynomials not only have an explicit expression, but are also expressed in terms of the same polynomials again, now evaluated at specific loci. The name superintegrability refers to a rather stretched analogy with classical mechanics, where some systems have extra integrals of motion which allows to reduce the problem to algebra and present an explicit solution.
		\\\\
		The superintegrability property is simplest in such models as the Hermitian Gaussian and complex matrix models \cite{Mironov:2017och}, but it is remarkable in the sense that it allows one to guess generalizations to other cases. One just needs to guess the appropriate substitute of polynomials. One such interesting generalization is the $\beta$-deformation \cite{Morozov:2012dz}. The corresponding matrix model is also referred to as the $\beta$-ensemble. Such $\beta$-deformed integrals are interesting in two ways. First, of all on their own they represent an eigenvalue model in which the quantum measure is deformed and a lot of familiar structures break down. This seems to be an important deformation direction as such models find an application in a variety of problems such as supersymmetric localization \cite{Cassia:2020uxy, Cassia:2019sjk}, categorification of knot invariants \cite{Dunin-Barkowski:2011ahc, Aganagic:2011sg, Gorsky:2013jna, Cherednik:2011nr}, AGT correspondence\cite{Awata:2010yy}. On the other hand completely understanding $\beta$-deformation is necessary to move the more general cases of cases like $(q,t)$ and elliptic $(q,t)$ matrix models \cite{Mironov:2020alu}.
		
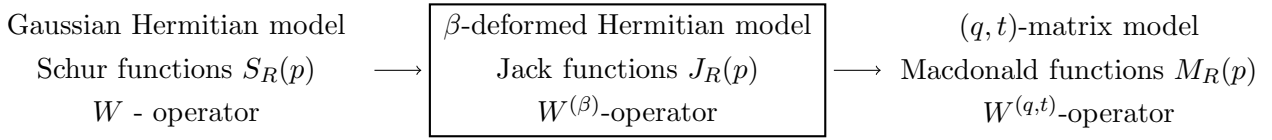
\vspace{0.5cm}
			\begin{center}
	    	\begin{tikzpicture}[scale=0.75]
	    	\node (HMM) at (-8,0) {
	    		\begin{minipage}{5cm}
		    	\begin{center}
			    	Gaussian Hermitian model\\
			    	\vspace{0.1cm}
				    Schur functions $ S_R(p) $ \\
				    	\vspace{0.1cm}
	    			$ W $ - operator
		    	\end{center}
			    \end{minipage}};
    		\node (beta) at (0,0) [rectangle,draw,thick]{
	    		\begin{minipage}{5cm}
		    	\begin{center}
			    	$\beta$-deformed Hermitian model\\
			    		\vspace{0.1cm}
				   Jack functions $ J_R(p) $ \\
				   	\vspace{0.1cm}
    				$ W^{(\beta)} $-operator
	    		\end{center}
		    	\end{minipage}}
			    ;
    		\node (qt) at (8,0) {
	    		\begin{minipage}{5cm}
		    	\begin{center}
			    	$(q,t)$-matrix model\\
			    		\vspace{0.1cm}
				    Macdonald functions $ M_R(p) $ \\
				    	\vspace{0.1cm}
    				$ W^{(q,t)}$-operator 
	    		\end{center}
		    	\end{minipage}};
		\draw[shorten >=0.1cm,->] (HMM)--(beta);
		\draw[shorten <=0.1cm,->] (beta)--(qt);
		\end{tikzpicture}
		\end{center}
\vspace{0.5cm}
		The appropriate symmetric functions for the $\beta$-deformed model are the so-called Jack polynomials \cite{macdonald1998symmetric}, for which one has \cite{Morozov:2019gbt,Cassia:2020uxy}:
		\begin{equation}
		\label{intojack}
			\langle J_{R}(H)\rangle= \beta^{|R|}\frac{J_R(N)}{J_R(\delta_{k,1})}\frac{J_R(\delta_{k,2})}{\Vert J_R\Vert^2 },
		\end{equation}
		which can be verified in several ways. However, there is still some mystery about proving such type of formula. In this paper we present a method of solving matrix models which naturally produces polynomial averages. We apply it to the $\beta$-deformed model,  thus providing the long-awaited proof of \eqref{intojack}
	    \\\\
		Other features of matrix models which are important in our discussion are ordinary KP/Toda integrability \cite{Morozov:1994hh}, the Virasoro constraints and $W$-representation \cite{Morozov:2009xk, Mironov:1990im}. These are matrix models analogues of conservation laws and equations of motions. Whether this analogy could be developed further is an intriguing question. Nevertheless, these three structures are crucial for our construction. 
		\\\\
		The former, is the idea that  matrix models partitions functions are resolvents of integrable systems. This is manifested in bilinear equations satisfied by matrix model partition functions. From the algebraic point of view it means that partition functions are certain matrix elements of the $GL(\infty)$ group. From this perspective it is natural to expect the character expansion of such matrix models to be in terms of $GL(\infty)$ characters - Schur functions \cite{date1982transformation}. 
		\\\\
		On the other hand there are Virasoro constraints (called Ward identities in QFT), which are linear differential equations, annihilating the partition function. They reflect the invariance of the integral under arbitrary reparametrizations of the integration variable and substitute the equations of motion for the path integral. Namely, the full set of Virasoro constraints completely determines the partition function. One could wish to be able to solve matrix models -  obtain full partition functions, by solving the Virasoro equations. Lately it was found that this is possible (at least up to choice of integration contours). The answer is typically given by the W-representation \cite{2021} -  an evolution  operator that generates the partition function from the trivial one:
		\begin{equation}\label{Wgen1}
			Z=e^{W}\cdot 1
		\end{equation}
		However, such answer is still unsatisfactory. Despite that one can write out the explicit $W$-operator, it is complicated enough that is not immediately clear how to expand this expression and obtain some explicit formulas. In this paper we explain how to promote
		\eqref{Wgen1} to an explicit expansion and hence incorporate superintegrability. 
	    The main idea is that the $W$-operator acts naturally on characters:
		\begin{equation}\label{Wgen2}
		W \chi_R = \sum_{R'} c_{RR'} \chi_{R'}
        \end{equation}
        where sums are restricted to additions of just a few boxes
        to the original Young diagram $R$ and coefficients $c_{RR'}$ are factorized contributions of some combinatoric piece  and the content factors $(i-j+N)$ or their proper deformations, with $(i,j)$ denoting the coordinates of these boxes. This property of $W$-operators is one manifestation of them being special elements of the $\mathcal{W}_{\infty}$ algebra.
        The example that we treat here demonstrates that these algebraic properties survive  $\beta$-deformation \cite{Morozov:2012dz}. On the other hand KP-like integrability seems to break down: there are no determinant formulae or bilinear identities and the substitute for $GL(\infty)$ is unknown. However, the $W$-representation is deformed nicely and formulae like \eqref{Wgen2} are still there with appropriate substitutes of characters \cite{Morozov:2019gbt}. 
		We will explain how the $\beta$-deformed $W$-operator acts on Jack polynomials and how this action allows us to calculate character averages \eqref{intojack} without any integration (see similar constructions in \cite{Wang:2021kwp,Kang:2021mqb}).
		\\\\
		We describe this method for the special case of the $\beta$-deformed Gaussian model, however, we keep in mind that is applicable in a number of other cases. Only minor adaptions are required in cases where contour ambiguities are absent and the relevant function are Schur functions or Jack polynomials. This includes the Gaussian model, the complex matrix model, the model with logarithmic potential and it's $\beta$-deformation. Furthermore, it is also applicable to the Kontsevich models, where the characters are Schur $Q$-functions.
		\\\\
		This paper is organized as follows. In the Section \ref{s2} we illustrate the proof of superintegrability on the example of undeformed Hermitian Gauss matrix model and Schur polynomials. Next, we prove superintegrability for the $\beta$-deformation in section \ref{s3}. The connection between Calogero-Sutherland Hamiltonians and $W$-operators is discussed in the section \ref{4}. Finally, we briefly discuss our result and further directions in section \ref{s5}.
		\\\\
    	At the moment of finalizing this paper we became aware that a very similar consideration has just appeared in a wonderful paper \cite{Wang:2022lzj}.

	\section{Undeformed case}\label{s2}
		Let us start with the standard Hermitian Gaussian matrix model and remind how one obtains explicit expression for averages in this case. The partition function of the model is:  
		\begin{equation} \label{eq24} 
			Z(p_k)=\int dH \exp\left(-\frac{1}{2} \Tr H^2 + \sum_k p_k \Tr H^k \right) 
		\end{equation}
		This model belongs to the class of so-called eigenvalue models, i.e. one can integrate out its angular part. After transition to eigenvalues $\lambda_i$ of $H$ the generating function becomes
		\begin{equation} \label{eq1} 
			Z(p_k)=\int d\lambda_1 ... d\lambda_N\, \Delta^{2} (\lambda) \exp\left[ -\frac{1}{2}\sum_{i}\lambda_i^2\right] \exp\left[ \sum_{k}p_k\sum_{i}\lambda_i^k\right] 
		\end{equation} 
		where $\Delta$ is the Vandermonde determinant: $ \Delta=\prod_{i<j}(\lambda_i-\lambda_j) $. 
		\\\\
An important property of the partition function is its expansion in terms of characters. Recall the Cauchy identity:
		\begin{equation}\label{eq36}
			\exp\left(\sum_{k} \dfrac{p_k \bar{p}_k}{k} \right) = \sum_R S_R(p_k)S_R(\bar{p}_k).
		\end{equation}
		Applying it to the potential of the matrix model one obtains:
		\begin{equation} \label{eq2} 
			Z=\sum_{R}S_{R}(p_k)\int dH \exp\left(-\frac{1}{2}\Tr H^2 \right) S_{R}(H)=\sum_{R}S_{R}(p_k)\left\langle S_{R}(H)\right\rangle 
		\end{equation}
		Hence, knowing all character  means we have an explicit perturbative solution of the matrix model. Clearly, since Schur polynomials form a basis in the space of all symmetric functions, we can calculate the expectation value of any other gauge invariant operator provided we know, how it expands in Schur operators.
		\\\\
		The partition function satisfies a set of differential equations called Virasoro constraints. These reflect the invariance of the integral under changes of the integration variables. To derive the constraints one changes the integration variables $\lambda_i \rightarrow \lambda_i+\epsilon \lambda_i^{n+1}$ and expand by powers of $\epsilon$ \cite{Morozov:1994hh,Mironov:1990im}. In the first order one gets equations:
		\begin{equation}\label{eq16}
			L_n Z=0 \quad \, n \geq -1
		\end{equation}
		where $L_n$ are Virasoro operators:
		\begin{equation} \label{eq40}
			L_n=\left(2N\frac{\partial}{\partial p_{n}}+\sum_{k=1}^{\infty}k p_k\frac{\partial}{\partial p_{k+n}}+\sum_{r=1}^{n-1}\frac{\partial^2}{\partial p_{r}\partial p_{n-r}}+N^2\delta_{n,0}+p_1N\delta_{n,-1}-\frac{\partial}{\partial p_{n+2}}\right)
		\end{equation}
		\\\\
		Before explaining how we suggest to solve the Virasoro constraints, let us shortly review a few rather traditional ways of solving the matrix model, namely obtaining explicit Schur averages. We do it because \emph{all of these methods break down} in the $\beta$-deformed case, which explains why proving \eqref{intojack} is not simple. The method explained later in this section survives the $\beta$-deformation.
		\begin{itemize}
		\item First of all, there is, of course, Wick's theorem \cite{Mironov:2017aqv}. The key idea is to represent arbitrary correlators in terms of the symmetric group:
		\begin{equation}
			\left\langle\prod_{i=1}^{m} H_{a_{i} \alpha_{i}} H^{b_{i} \beta_{i}}\right\rangle=\sum_{\substack{\gamma \in S_{m} \\ [\gamma] = [2^m]}} \prod_{i=1}^{m} \delta_{a_{i}}^{b_{\gamma(i)}} \delta_{\alpha_{i}}^{\beta_{\gamma(i)}}
		\end{equation} 
		where the sum goes over permutations $\gamma$ with a fixed cycle type $[2^m]$, which then allows to represent the correlator of monomials in terms of symmetric group characters:
		\begin{equation}
			\left\langle\prod_{p=1}^{l_{\Lambda}}\operatorname{Tr}H^{m_{p}}\right\rangle =\sum_{R \vdash m} \varphi_{R}\left(\left[2^{m}\right]\right) \cdot D_{R}(N) \cdot \psi_{R}(\sigma)
		\end{equation}
		Here $\psi_R (\sigma)$ and $\varphi_R (\sigma)$ are differently normalized symmetric group characters for representation $R$ and cycle type $\sigma$, while $D_R(N)$ is the dimension of the corresponding $GL(N)$ representation and is equal to $S_R(N)$.\\\\
As  described in \cite{Mironov:2017aqv} one can use symmetric group character orthogonality to further construct Schur averages and explicitly obtain formula
		\begin{equation}\label{schursuperint}
			\langle S_{R}(H)\rangle=\frac{S_R(N)}{S_R(\delta_{k,1})} S_R(\delta_{k,2})
		\end{equation}
		\item  On the other hand, one can just do explicit angular integration over the unitary group \cite{Kazakov:1995ae}. Consider the integral:
		\begin{multline}\label{eq41}
			Z(Y)=\int dH \exp\left(-\frac{1}{2} \Tr H^2 + \Tr HY \right)=\\ = \int dH \exp\left(-\frac{1}{2} \Tr H^2 \right) \int[dU]\exp\left(\Tr UHU^{\dagger}Y \right)=\\
			= \sum_{|R|\leqslant N}\frac{S_R(\delta_{k,1})S_R(Y)}{S_{R}(N)}\int dH \exp\left(-\frac{1}{2} \Tr H^2 \right) S_R(H)=\sum_{|R|\leqslant N} \frac{S_R(\delta_{k,1})S_R(Y)}{S_{R}(N)}\left\langle S_R(H) \right \rangle.
		\end{multline}
		Here the transition between the second and third line is the character expansion of the Itzyckson-Zuber integral \cite{Morozov:2009jv}. Besides one can just take the Gaussian integral and apply the Cauchy formula:
		\begin{equation}\label{eq42}
			Z(Y)=\int dH \exp\left(-\frac{1}{2} \Tr H^2 + \Tr HY \right)=e^{\frac{1}{2}\Tr Y^2}= \sum_{R}S_R(Y)S_R(\delta_{k,2})
		\end{equation}
Comparing the two expression we immediately obtain \eqref{schursuperint}.
		\item 	Lastly, we could use integrability properties of the partition function \cite{Mironov:2021udg}. Namely, one can represent the partition function as a determinant of the moment matrix given by
		\begin{equation}
			\operatorname{\textswab{M}}_{n}\left\{p_{k}\right\}:=\int \exp \left(\sum_{k=1}^{\infty} \frac{p_{k} x^{k}}{k}\right) \cdot x^{n} \rho(x) d x
		\end{equation}
In terms of averages of Schur functions this means:
		\begin{equation}\label{cR}
	 	c_{R}:= \left\langle S_R \right \rangle	=\frac{\operatorname{det}_{1 \leq i, j \leq N} \operatorname{\textswab{M}}_{N-i+j+R_{i}-1}\{0\}}{\operatorname{det}_{1 \leq i, j \leq N} \operatorname{\textswab{M}}_{N-i+j-1}\{0\}}
		\end{equation}
		According to the general idea, mentioned in the introduction, we could use the lowest Virasoro constraint $L_{-1}$, also called the string equation. In terms of \eqref{cR} it is written as
		\begin{equation}
			\sum_{\square} c_{R+\square}=\sum_{\square}\left(N-i_{\square}+j_{\square}\right) c_{R-\square}.
		\end{equation}
		Solving for symmetric representations we determine the moments:
		\begin{equation}
			\mathrm{\textswab{M}}_{2 r}=(2 r-1) ! ! \cdot \operatorname{\textswab{M}}_{0} \quad \operatorname{\textswab{M}}_{2 r-1}=0
		\end{equation}
		which correctly reproduces the moments of the Gaussian measure. Finally, inserting the moments back into \eqref{cR} and after a few algebraic manipulations, which are explained in detail in \cite{2021}, we obtain \eqref{schursuperint}.
		
		\end{itemize}
		The idea of this paper is to solve the system of equations, written above, explicitly.  It turns out, that the system (\ref{eq16}) is equivalent to a single equation \cite{Mironov:2021udg}, which is the sum of Virasoro constraints:
		\begin{equation}\label{eq11}
			\sum_{n\geq 1} p_n L_{n-2}Z=0 
		\end{equation}
		The last equation can be written as 
		\begin{equation}\label{eq12}
			(l_0-2 W_{-2})Z(t)=0  
		\end{equation}
		where the operator $W_{-2}$ has degree\footnote{For monomial operator $ \prod_{i}p_i^{m_i}\prod_{j}\frac{\partial^{\alpha_j}}{\partial p_j^{\alpha_j}} $ degree is defined as $ \sum_{i}i m_i-\sum_{j}j \alpha_j $} 2 and  $ l_0=\sum np_n\frac{\partial}{\partial p_n} $ is nothing but the grading operator. This equation is special case of the equation
		\begin{equation}\label{eq13}
			(l_0-k\widehat{O}^{(k)})\Psi=0
		\end{equation}
		where $ \widehat{O}^{(k)} $ is an operator with degree $k$, i.e. $\left[ l_0, \hat{O}^{(k)} \right] =k \hat{O}^{(k)}$. The solution of the equation is 
		\begin{equation}\label{eq14}
			\Psi=e^{\widehat{O}^{(k)}} \cdot 1 
		\end{equation}
		Therefore, we obtain the partition function of the Hermitian matrix model:
		\begin{equation} \label{eq37}
			Z= e^{W_{-2}} \cdot 1
		\end{equation}
		Not only does one have an explicit solution, but can naturally recover the character expansion. For this, notice, that Schur polynomials are exactly "natural" for W-operator to act on. It acts on them by adding two boxes to the representation with some weight:
		\begin{equation}\label{eq10}
			W_{-2}S_R=\frac{1}{2}\sum_{R'=R+\square_1+\square_2}(j_{\square_1} -i_{\square_1}+N) (j_{\square_2}- i_{\square_2}+N) C_{RR'}S_{R'}
		\end{equation}
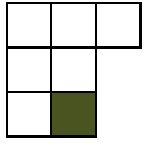
		\begin{center}
			\ydiagram{3,2,2}*[*(armygreen)]{0,0,1+1}\\
		\end{center}
		where $i_1$, $i_2$, $j_1$ and $j_2$ are coordinates of the positions of the boxes added to the initial diagram (in the picture painted box has coordinates $(i,j)=(2,1)$). The quantity $i-j$ is sometimes called the \emph{content} of the box in a Young diagram.	Coefficients  $ C_{RR'} $ come from the expansion $ p_2 S_R $ by Schur polynomials: 
		\begin{equation}\label{p2schur}
			p_2 S_R=\sum_{R'=R+\square+\square}C_{RR'}S_{R'} 
		\end{equation}			 
		and in this case vanish except if $R'$ differs from $R$ by a piece of form $[2]$ or $[1,1]$, in other word the skew-diagram $R'/R$ is a horizontal or vertical strip of size 2, and then $C_{RR'}=\pm 1$.
		For example:
		$$ p_2 S_{[2]}=S_{[4]}+0 \cdot S_{[3,1]}+ S_{[2,2]}-S_{[2,1,1]} $$
		$$ W_{-2}S_{[2]}=\frac{1}{2}\left[ (2+N)(3+N)S_{[4]}+0 \cdot S_{[3,1]}+(N-1)N S_{[2,2]}-(N-2)(N-1)S_{[2,1,1]} \right] $$		
		Thus it is convenient to rewrite (\ref{eq37}) in terms of Schur functions:
		\begin{equation}\label{eq15}
			Z=e^{W_{-2}}\cdot S_{\varnothing}(p_k)
		\end{equation}
		where $ S_{\varnothing}(p_k)=1 $ is Schur function of empty Young diagram. Thus knowing the expression (\ref{eq10}), by acting iteratively on Schur functions,  one obtains the character expansion of the partition function. From the ($\ref{eq15}$) one can extract superintegrability for Schur functions \eqref{schursuperint}. We will describe the procedure in more detail in the next section for the Jack functions straight away.		
		
	\section{$ \beta $-deformation} \label{s3}
		Now we are ready to present the main result of the paper. Namely, we prove the superintegrability of the $\beta$-deformation of the Gaussian Hermitian matrix model.
		\\\\
		This deformation  is introduced in the form of integral of eigenvalues (for a matrix integral representation of this model see \cite{tridiag,Mironov:2021bbx}). In general one can consider integrals not only of Hermitian matrices but of orthogonal or simplectic. The eigenvalue representation for both of these models will differ from the Hermitian in the power of the Vandermonde determinant in (\ref{eq1}), one for orthogonal and four for symplectic respectively. Hence it is only natural to study eigenvalue integrals with the power of the determinant being a parameter, now taking any value. Hence, we define the following partition function:
		\begin{equation} \label{eq26}
			Z_\beta(p_k)=\int d\lambda_1 ... d\lambda_N \, \Delta^{2\beta}(\lambda) \exp\left[ -\frac{1}{2}\sum_{i}\lambda_i^2\right]\exp\left[ \sum_{k}\beta p_k\sum_{i}\lambda_i^k\right] 
		\end{equation}\\
		One can still expand the partition function in terms of characters. However, it is now well known that in this case the proper basis functions are so-called Jack polynomials. They are symmetric polynomials orthogonal with respect to a certain scalar product, and reduce to Schur functions for $\beta=1$ \cite{macdonald1998symmetric}. We list some examples of the simplest Jack polynomials for illustration: 
			\begin{equation}
		\begin{split}
		&J_{[1]}=p_1 \\
		&J_{[2]}=\frac{1}{\beta+1}(\beta p_1^2+p_2) \\
		&J_{[1,1]}=\frac{1}{2}(p_1^2-p_2) 
		\end{split}
		 \qquad \qquad
			\begin{split}
			&J_{[3]}=\frac{1}{(\beta+1)(\beta+2)}(\beta^2 p_1^3+3\beta p_1 p_2+2p_3)\\
			&J_{[2,1]}=\frac{1}{2\beta+1}(\beta p_1^3+(1-\beta)p_1 p_2-p_3)\\
			&J_{[1,1,1]}=\frac{1}{6}(p_1^3-3p_1p_2+2p_3)
			\end{split}
			\end{equation}
	The Cauchy identity for has a $\beta$-deformation as well, therefore for Jack polynomials we have:
		\begin{equation}\label{eq25}
			\sum_R\frac{x^{|R|}}{\Vert J_R\Vert^2}J_R(p)J_R(\bar{p})=\exp\left[ \sum_k \beta  x^k  \frac{p_k \bar{p}_k}{k}\right]
		\end{equation}
		where $||J_R||^2$ is the norm of the Jack polynomial. By proper functions we mean that averages of Jack polynomials constitute a direct $\beta$-deformation of (\ref{schursuperint}). The expectation value of Jack polynomials in the $\beta$-deformed Hermitian Gaussian model is given by
		\begin{equation} \label{eq9}
			\langle J_{R}(H)\rangle=\frac{J_R(N)}{J_R(\delta_{k,1})}\frac{J_R(\delta_{k,2})}{\Vert J_R\Vert^2 }\beta^{|R|}
		\end{equation}
		\\
		A key difference is that, as we have mentioned, it seems harder to prove this formula. Clearly, we cannot efficiently use Wick's theorem. Standard KP/Toda integrability breaks down, i.e. no determinant-like representation in known for the partition function. However, we can still solve the model using Virasoro constrains. The Virasoro and W-operators are obtained the same way as in the undeformed case:
		\begin{multline} \label{eq4}
			L^{(\beta)}_n=((n+1)(1-\beta)+2N\beta)\frac{\partial}{\partial p_{n}}+\beta \sum_{k=1}^{\infty}k p_k\frac{\partial}{\partial p_{k+n}}+\beta^2 \sum_{r=1}^{n-1}\frac{\partial^2}{\partial p_{r}\partial p_{n-r}}+\\
			+((1-\beta)+N\beta)\beta N\delta_{n,0}+p_1 \beta^2 N\delta_{n,-1}-\frac{\partial}{\partial p_{n+2}}
		\end{multline}
		rewriting
		\begin{equation}\label{eq17}
			\sum_{n\geq 1} p_n L^{(\beta)}_{n-2}Z_\beta(p_k)=0 
		\end{equation}
		as 
		\begin{equation}\label{eq18}
			(l_0-2 W_{-2}^{(\beta)})Z_{\beta}(p_k)=0  
		\end{equation}
		one obtains $\beta$-deformed W-operator:
		\begin{multline} \label{eq5}
			 W_{-2}^{(\beta)}=\sum_{n=1}^{\infty} \left(\frac{(n+1) (1-\beta)}{2}+N\beta\right) p_{n+2} \frac{\partial}{\partial p_{n}} +\frac{\beta}{2} \sum_{k,n=1}^{\infty} (n+k-2)k p_{n}p_k \frac{\partial}{\partial p_{k+n-2}} +\\
			 +\frac{1}{2}\sum_{k,n=1}^{\infty}n k p_{k+n+2}\frac{\partial^2}{\partial p_{k}\partial p_{n}}+\frac{((1-\beta)+N\beta)}{2}\beta N p_2+\frac{1}{2} \beta^2 p_1^2N
		\end{multline}\\
		It turns out that Jack polynomials are "natural" functions for $\beta$-deformed W-operator too:
		\begin{equation}\label{eq6}
			W_{-2}^{(\beta)}J_R=\frac{1}{2}\sum_{R'=R+\square_1+\square_2}(j_{\square_1}+\beta(N-i_{\square_1})) (j_{\square_2} +\beta(N-i_{\square_2}))C_{RR'}J_{R'}
		\end{equation}
		where $ C_{RR'} $ are coefficients of expansion $ p_2 J_R $ in terms of Jack polynomials:
		\begin{equation} \label{eq19}
			p_2 J_R=\sum_{R'=R+\square+\square}C_{RR'}J_{R'}
		\end{equation}
		As in the underformed case, the important property is that the action of the $W$-operator differs only by a box content factor. For example:
			\begin{equation}\label{eq38}
				\begin{split}
				 p_2 J_{[2]}&=
				 \\
				=&J_{[4]}-\frac{2(\beta-1)\beta}{(\beta+1)(\beta+3)} J_{[3,1]}+\frac{4(1+2\beta)}{(1+\beta)^2(2+\beta)} J_{[2,2]}-\frac{2\beta(1+3\beta)}{(1+\beta)^3}J_{[2,1,1]} 
				\end{split}
			\end{equation}	
			while 
			\begin{equation}\label{eq39}
				\begin{split}		
				W_{-2}^{(\beta)}J_{[2]}&=
				\\
				=&\frac{1}{2}[(2+N\beta)(3+N\beta)J_{[4]}-\beta(N-1)(N\beta+2)\frac{2(\beta-1)\beta}{(\beta+1)(\beta+3)} J_{[3,1]}+\\ 
		 		&+\beta(N-1)((N-1)\beta+1)\frac{4(1+2\beta)}{(1+\beta)^2(2+\beta)} J_{[2,2]}-\beta(N-2)\beta(N-1)\frac{2\beta(1+3\beta)}{(1+\beta)^3}J_{[2,1,1]}
				\end{split}
			\end{equation}		 
			Now we would like to describe in detail how an iterative application of formula \eqref{eq6} leads to an explicit expression for expectation values of Jack polynomials. After iterated application of (\ref{eq6}) and (\ref{eq19}) one can obtain		
			\begin{equation} \label{eq20}
				(W_{-2}^{(\beta)})^{n}\cdot J_{\varnothing}=\frac{1}{2^n}\sum_{R_n:|R_n|=2n}\left(\prod_{ \left(i_\square,j_\square \right) \in R_n}  \left( j_\square+\beta(N-i_\square)\right)\right) \sum_{\left\{ {R_1,\ldots ,R_{n-1}} \right\} }D_{\varnothing,R_1,R_2,...,R_n}J_{R_n}
			\end{equation}
			\begin{equation} \label{eq21}
				p_2^n \cdot J_{\varnothing}=\sum_{R_n:|R_n|=2n}\sum_{\left\{ {R_1,\ldots ,R_{n-1}} \right\} }D_{\varnothing,R_1,R_2,...,R_n}J_{R_n}
			\end{equation}
			Here $ D_{\varnothing,R_1,R_2,...,R_n}=C_{\varnothing,R_1}C_{R_1,R_2}...C_{R_{n-1},R_n} $; $ R_i: |R_i|+2=|R_{i+1}| $ are combinatorial coefficients, which correspond to a certain pattern in which one obtains the Young diagram $R$ from an empty one. The sum is taken over all such sequences $\left\{ {R_1,\ldots ,R_{n-1}} \right\}$ in which every next partition is obtained from the previous one by adding two boxes according to the coefficient $C_{R_1,R_2}$.  In other words, it is a sum over "paths" in the set of Young diagrams, where each "path" comes with a certain weight governed by formula \eqref{eq21}.
			\\\\
			As an example, representation $[3,1]$ can be obtained in two ways (they are illustrated on the image below): 1) on the first step adding boxes with coordinates $(0,0)$ and $(0,1)$, on the second step - $(0,2)$ and (1,0); 2) on the first step adding boxes with coordinates (0,0) and (1,0), on the second step - (0,1) and (0,2). Paths $ D_{\varnothing,[2],[3,1]}$ and $ D_{\varnothing,[1,1],[3,1]}$ correspond to these ways respectively. 
			
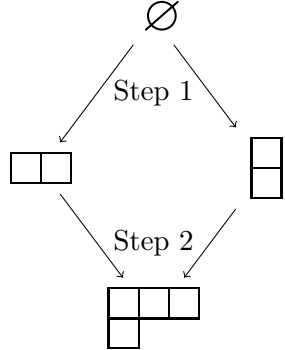
\begin{center}
		\begin{tikzpicture}[scale=0.5]
		\ytableausetup{boxsize=1em,aligntableaux = center}
		\node (empty) at (0,0) {
			\scalebox{1.75}{
			\begin{minipage}{3cm}
			\begin{center}
				$\varnothing$
			\end{center}
			\end{minipage}}
			};
		\node (2) at (-3,-4) {
			\begin{minipage}{3cm}
			\begin{center}
				\ydiagram{2}
			\end{center}
			\end{minipage}};
		\node (11) at (3,-4) {
			\begin{minipage}{3cm}
			\begin{center}
				\ydiagram{1,1}
			\end{center}
			\end{minipage}};
		\node (step1) at (0,-2) {
			\begin{minipage}{3cm}
			\begin{center}
				Step 1
			\end{center}
			\end{minipage}};
		\node (step2) at (0,-6) {
			\begin{minipage}{3cm}
			\begin{center}
				Step 2
			\end{center}
			\end{minipage}};
		\node (31) at (0,-8) {
			\begin{minipage}{3cm}
			\begin{center}
				\ydiagram{3,1}
			\end{center}
			\end{minipage}};
		\node (left) at (-2,0){};
		\node (right) at (2,0){};
		\draw[->] (empty)--(2);
		\draw[->] (empty)--(11);
		\draw[<-] (31)--(2);
		\draw[->] (11)--(31);
		\end{tikzpicture}
		\end{center}
			A key observation is that the piece $ \prod\limits_{(i_\square,j_\square) \in R_n} (j_\square + \beta (N- i_\square )) $ factors out from the sum over "paths" because it does not depend on the order in which each box is added, but only on the content.The expression for this factor is a version of the hook-content product formula, see \cite{macdonald1998symmetric} formula (10.25):
			\begin{equation} \label{eq7}
				\frac{J_R(N)}{J_R(\delta_{k,1})}=\beta^{-|R|} \prod_{ \left(i_\square,j_\square \right)  \in R} (j_\square+\beta(N-i_\square))
			\end{equation}
			A typical example is:
			$$ \frac{J_{[3,2]}(N)}{J_{[3,2]}(\delta_{k,1})}=\frac{\beta(M-1) \beta N (N\beta+1)(N\beta+2)(1+\beta(N-1))}{\beta^5}. $$
		Now, lets return to the evaluation of the combinatorial sum. We don't need to know each term. The trick here is to use the fact that this sum originates in the Pieri-like formula \eqref{eq21}.	From the Cauchy identity follows 
		$$ e^{p_2} =\sum_{R}\frac{J_R(p_k) J_R(\delta_{k,2})} {\Vert J_R\Vert^2} 2^{|R|/2} $$
		Rewriting it in the same manner as the iterative action of the $W$-operator and using relation (\ref{eq21}) we obtain:
			\begin{equation}\label{eq22}
		 		\sum_{R}\frac{J_R(p_k)J_R(\delta_{k,2})}{\Vert J_R\Vert^2}2^{|R|/2}=e^{p_2} \cdot 1=\sum_{R:|R|-even}J_R(p_k)\sum_{\left\{ {R_1,\ldots ,R_{n-1}} \right\} }D_{\varnothing,R_1,R_2,...,R_n}\frac{1}{(|R|/2)!} 
		 	\end{equation}
			Thus we obtain
			\begin{equation}\label{eq23}
		 		\frac{\Vert J_R\Vert^2}{2^{|R|/2}(|R|/2)!}\sum_{\left\{ {R_1,\ldots ,R_{n-1}} \right\} }D_{\varnothing,R_1,R_2,...,R_n} = J_R(\delta_{k,2})
		 	\end{equation}
		 	Let us illustrate how this works:
			$$ \frac{J_{[3,1]}(\delta_{k,2})}{\Vert J_{[3,1]}\Vert^2}=\frac{1}{2}[D_{\varnothing,[2],[3,1]}+D_{\varnothing,[1,1],[3,1]}]=\frac{1}{2}[C_{\varnothing,[2]}C_{[2],[3,1]}+C_{\varnothing,[1,1]}C_{[1,1],[3,1]}]=$$
			$$=\frac{1}{2}\left[-\frac{2\beta(\beta-1)}{(\beta+1)(\beta+3)}-\frac{2\beta}{1+\beta}\right]=-\frac{2\beta}{\beta+3} $$
			Hence, using formula (\ref{eq23}) we obtain the partition function:
			\begin{multline}\label{eq8}
		 		Z_\beta=e^{W_{-2}^{(\beta)}}\cdot J_{\varnothing}(p_k)=\sum_{R:|R|-even}J_R(p_k)\frac{1}{(|R|/2)!}\sum \limits_{\left\{ {R_1,\ldots ,R_{n-1}} \right\}} D_{\varnothing,R_1,R_2,...,R_n} \prod \limits_{(i_\square,j_\square) \in R_n}(j_\square+\beta(N-i_\square))=\\
		 		=\sum_{R:|R|-even}J_R(p_k)\frac{J_R(N)}{J_R(\delta_{k,1})}\frac{J_R(\delta_{k,2})\beta^{|R|}}{\Vert J_R\Vert^2 } 
		 	\end{multline}
	\section{Constructing $W$-operators from Hamiltonians}\label{4}
		Formula \eqref{eq6} is the key to our construction, however we did not provide an explicit proof. Here we are going to sketch a general idea of where such operators come from and how to prove they act on characters in the mentioned way. The construction is rather similar to the one considered in \cite{lassalle2009jack} and \cite{Awata2000}. We postpone a complete analysis, which would also involve the $(q,t)$-deformed case to a separate paper. 
		\\\\
		Suppose instead of $W_{-2}^{(\beta)}$ \eqref{eq6} we got a simpler operator $W_{-1}^{(\beta)}$, which acts on Jack polynomials as
		\begin{equation} \label{eq31}
			W_{-1}^{(\beta)}J_R=\sum_{R'=R+\square}(\beta(N-i_{\square})+j_{\square})C_{RR'}J_{R'}
		\end{equation}
		There $C_{RR'}$ are the coefficients of expanding $p_1J_R$ over Jack polynomials:
		\begin{equation}\label{eq27}
			p_1J_R=\sum_{R'=R+\square} C_{RR'}J_{R'}
		\end{equation}
		To prove \eqref{eq31} we notice that it can be constructed by commuting the multiplication operator $p_1$ with a diagonal operator:
		\begin{equation}
			H^{(\beta)}_{1} J_R = \sum_{( i_\square, j_\square) \in R} \left( \beta \left(N-i_\square \right) + j_\square \right) J_R
		\end{equation}		
		Such operators, diagonal in the Jack polynomial basis, are nothing but Calogero-Ruijsenaars Hamiltonians. We need their expression in terms of time variables. In \cite{Awata2000, 2013, Zenkevich:2014lca} there is description of these operators, but for our current goal only one of them is needed. In our normalisation it reads
		\begin{equation} \label{eq28}
			H^{(\beta)}_1=\dfrac{1}{2}\sum_{n,m\geq 1}\left(n m p_{n+m}\frac{\partial^2}{\partial p_n\partial p_m}+\beta(n+m)p_n p_m\frac{\partial}{\partial p_{n+m}}\right)+ \dfrac{1}{2}\sum_{n\geq 1}((n+1)(1-\beta)-2\beta N)n p_n\frac{\partial}{\partial p_n}
		\end{equation}
		Finally, we can find the expression for $W_{-1}^{(\beta)}$ in terms if time variables:
	\begin{equation}\label{W1comm}
	    \begin{split}
	   	W_{-1}^{(\beta)}=[H^{(\beta)}_1,p_1] = \sum_{n}n p_{n+1}\frac{\partial}{\partial p_{n}}+p_1( 1-\beta - \beta N )
	    \end{split}
	\end{equation}	
		This procedure can be generalized and applied to proving relations similar to (\ref{eq31}). In particular, to prove (\ref{eq6}) one should construct $W_2$ from $H_2$ and $p_2$ in addition to $H_1$ and $p_1$. 
		
\section{Discussion}\label{s5}
The main technical result of this paper is the proof of formula \eqref{intojack}, expressing averages of Jack polynomials in terms of the same functions evaluated at special points.  At a more conceptual level we have developed a method of solving Virasoro equations explicitly in terms of the $W$-representation, which is applicable when usual ways of integration do not work. The algebraic side of the picture involves a representation of the $W$-operator in the space of characters. As we see the construction survives the $\beta$-deformation. From the discussion in section \ref{4} it is clear that there should be an immediate generalization to the case of $(q,t)$-deformation with the appropriate Macdonald Hamiltonians and further to elliptic models and possibly even further involving Kerov functions \cite{Mironov:2019uaa}(or non-Kerov deformations of Macdonald polynomials \cite{Awata:2020xfq}). As we can see out of all possible operators of the form \eqref{W1comm} matrix models select some specific ones. It would be interesting to distinguish matrix models out of "all models"  from this point of view.
\\\\
A lot of other intriguing directions of generalization immediately come to mind. First is the case of non-gaussian models and in general models with boundary conditions or non-trivial contour choices. It is distinguished by the fact that Virasoro constraints are not enough to fully specify the partition function, hence it seems that something should break in our method. On the other hand for specific choices of integration contours or boundary conditions superintegrabilty still holds and we could expect some kind of $W$-representation too \cite{Cordova:2016jlu,Cassia:2021dpd}.   
The second interesting direction is the generalized Kontsevich model. Here the situation is the opposite. The correct generalization of $W$-operators is known \cite{Mironov:2021euq}, however the appropriate characters are not (for attempts, see \cite{Mironov:2021lbx}). It seems like to obtain an answer in this case a deeper understanding of the relation between the algebra of $W$-operators and characters is needed. 

\section*{Acknowledgements}
We are grateful to A. Mironov, A. Morozov and A. Popolitov for insightful comments and stimulating discussions. We also thank A. Sleptsov for reading the paper and providing valuable comments.
\\
This work is supported by the Russian Science Foundation (Grant No.21-12-00400).

\bibliographystyle{utphys}
\bibliography{superint_beta}{}
	\end{document}